%% file: main.tex
\documentclass[conference]{IEEEtran}

\input{preamble.tex}

\title{Transformer-Based Speech Synthesizer\\ Attribution in an Open Set Scenario}

\author{
    \IEEEauthorblockN{\textbf{Emily R. Bartusiak and Edward J. Delp}}
    \IEEEauthorblockA{Video and Image Processing Lab, School of Electrical and Computer Engineering, Purdue University, West Lafayette, IN} 
}

\begin{document}
\maketitle

\begin{abstract}
Speech synthesis methods can create realistic-sounding speech, which may be used for fraud, spoofing, and misinformation campaigns.
Forensic methods that detect synthesized speech are important for protection against such attacks.
Forensic attribution methods provide even more information about the nature of synthesized speech signals because they identify the specific speech synthesis method (\ie speech synthesizer) used to create a speech signal.
Due to the increasing number of realistic-sounding speech synthesizers, we propose a speech attribution method that generalizes to new synthesizers not seen during training.
To do so, we investigate speech synthesizer attribution in both a closed set scenario and an open set scenario.
In other words, we consider some speech synthesizers to be ``known" synthesizers (\ie part of the closed set) and others to be ``unknown" synthesizers (\ie part of the open set).
We represent speech signals as spectrograms and train our proposed method, known as compact attribution transformer (CAT), on the closed set for multi-class classification.
Then, we extend our analysis to the open set to attribute synthesized speech signals to both known and unknown synthesizers.
We utilize a t-distributed stochastic neighbor embedding (tSNE) on the latent space of the trained CAT to differentiate between each unknown synthesizer.
Additionally, we explore poly-1 loss formulations to improve attribution results.
Our proposed approach successfully attributes synthesized speech signals to their respective speech synthesizers in both closed and open set scenarios.
\end{abstract}


\input{part-1-introduction}
\input{part-2-method}

\input{part-3-results}
\input{part-4-conclusion}

\section*{Acknowledgments}
This paper is based on research sponsored by the Defense Advanced Research Projects Agency (DARPA) and the Air Force Research Laboratory (AFRL) under agreement number FA8750-20-2-1004. The U.S. Government is authorized to reproduce and distribute reprints for Governmental purposes notwithstanding any copyright notation thereon. The views and conclusions contained herein are those of the authors and should not be interpreted as necessarily representing the official policies or endorsements, either expressed or implied, of DARPA, AFRL, or the U.S. Government.\\
\\
Address all correspondence to Edward J. Delp at \url{ace@ecn.purdue.edu}. 

\bibliographystyle{IEEEtran}
\bibliography{refs}

\end{document}

%% file: preamble.tex
\usepackage{cite}
\usepackage{amsmath,amssymb,amsfonts}
\usepackage{algorithmic}
\usepackage{graphicx}
\usepackage{textcomp}
\usepackage[table]{xcolor}  
\usepackage{booktabs, makecell, booktabs,multirow} 
\usepackage{url} 
\usepackage{hyperref}
\usepackage{subcaption}
\usepackage{flushend}

\def\BibTeX{{\rm B\kern-.05em{\sc i\kern-.025em b}\kern-.08em
    T\kern-.1667em\lower.7ex\hbox{E}\kern-.125emX}}
    
\newcommand{\etal}{\textit{et al}.\@ }
\newcommand{\ie}{\textit{i.e.},\@ }
\newcommand{\eg}{\textit{e.g.},\@ }

\newcolumntype{L}[1]{>{\raggedright\let\newline\\\arraybackslash\hspace{0pt}}m{#1}}
\newcolumntype{C}[1]{>{\centering\let\newline\\\arraybackslash\hspace{0pt}}m{#1}}
\newcolumntype{R}[1]{>{\raggedleft\let\newline\\\arraybackslash\hspace{0pt}}m{#1}}

\definecolor{kellygreen}{rgb}{0.3, 0.73, 0.09}
\definecolor{amber}{rgb}{1.0, 0.75, 0.0}
\definecolor{amethyst}{rgb}{0.6, 0.4, 0.8}
\definecolor{ballblue}{rgb}{0.13, 0.67, 0.8}
\definecolor{babyblue}{rgb}{0.54, 0.81, 0.94}
\definecolor{babyblueeyes}{rgb}{0.63, 0.79, 0.95}
\definecolor{bondiblue}{rgb}{0.0, 0.58, 0.71}
\definecolor{brightcerulean}{rgb}{0.11, 0.67, 0.84}
\definecolor{celestialblue}{rgb}{0.29, 0.59, 0.82}
\definecolor{cornflowerblue}{rgb}{0.39, 0.58, 0.93}
\definecolor{cyan(process)}{rgb}{0.0, 0.72, 0.92}
\definecolor{darkturquoise}{rgb}{0.0, 0.81, 0.82}
\definecolor{hanblue}{rgb}{0.27, 0.42, 0.81}
\definecolor{iris}{rgb}{0.35, 0.31, 0.81}
\definecolor{lightskyblue}{rgb}{0.53, 0.81, 0.98}

%% file: part-1-introduction.tex
\section{Introduction}\label{part-1-intro}

Many free, easy-to-use tools offer text-to-speech or voice conversion utilities to generate synthetic speech~\cite{kim2021conditional, wang2021prosody, popov2021gradtts, zhou2021seen, hutchinson_2021, resemble_2021}.
Synthesized speech can provide many benefits to society, such as accessibility services to support people who are visually impaired. 
However, synthesized speech can also be used for malicious purposes.
In 2021, an impersonator using synthesized speech on a conference call tried to convince Goldman Sachs to make a \$40 million investment~\cite{smith_2021}.
In 2022, a deepfake video showed Ukrainian President Volodymyr Zelensky surrendering to Russia~\cite{allyn_2022}.
A recent AI-produced podcast contains a realistic-sounding conversation between Steve Jobs and Joe Rogan, but all of the speech is synthesized~\cite{rogan_2022}.
These examples highlight the importance of audio forensics.
Large amounts of synthesized speech may be generated for large-scale scams and disinformation campaigns.
The same speech synthesizers are often utilized throughout the scams.
Thus, identifying the speech synthesizer used to create speech signals in these campaigns can reveal more information about how the scams are spreading and (possibly) who created them. 

In this paper, we investigate speech synthesizer attribution, which is the task of identifying which speech synthesizer was used to generate a synthetic speech signal.
To attribute a speech signal to a speech synthesizer, we convert speech signals into spectrograms\cite{rs2010}. 
A spectrogram is a 2-D temporal-spectral representation of a speech signal.
We treat the spectrograms as images and analyze them with our proposed deep learning method, called compact attribution transformer (CAT). 
We consider eleven different speech synthesizers in the scope of this work.
Eight of the synthesizers are used to train CAT.
The remaining three synthesizers are only used during testing to represent a set of unknown speech synthesizers that have not been seen during training.
We evaluate our method on both a closed set of known speech synthesizers as well as an open set, which contains a combination of known and unknown synthesizers.
We demonstrate that our proposed approach successfully attributes synthesized speech signals to their synthesizers and can differentiate between known and unknown synthesizers.
We utilize a t-distributed stochastic neighbor embedding (tSNE)~\cite{Maaten_2008} to separate synthesized speech signals created by different synthesizers in CAT's latent space.
We show that tSNE successfully discriminates different synthesizers, including different unknown synthesizers.
Thus, our proposed approach generalizes to new synthesizers.
Finally, we improve attribution performance by using poly-1 loss formulations.

\section{Related Work}\label{related-work}

\begin{figure*}[ht]
    \centering
    \includegraphics[width=.8\textwidth]{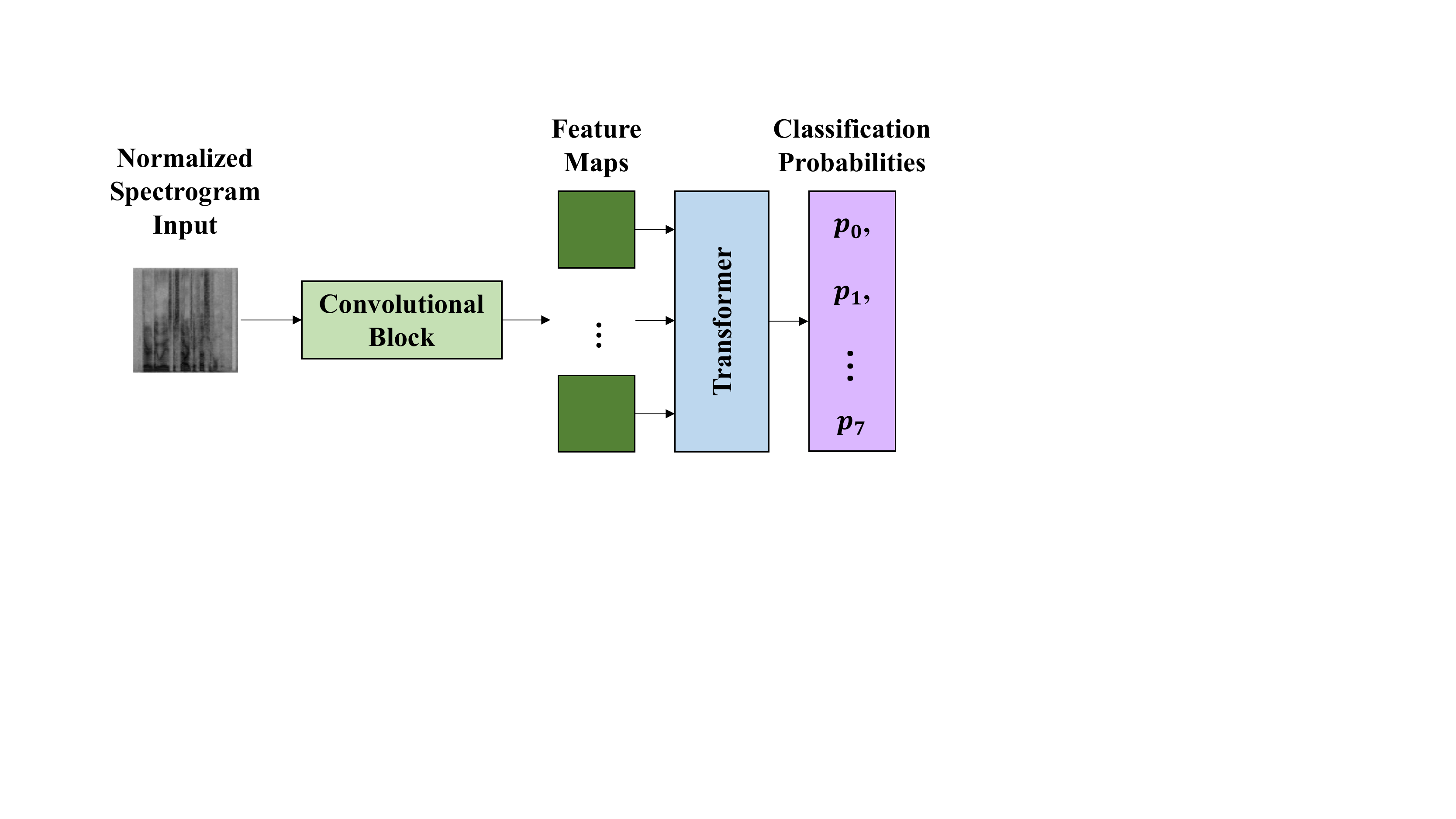}
    \caption{Block diagram of our proposed approach, known as compact attribution transformer (CAT).}
    \label{fig:overview}
    \vspace{-.25cm}
\end{figure*}

Most audio forensics methods focus on detecting manipulated or synthesized speech~\cite{long-mipr, bartusiak_2021_ei, hao_2022_deepfake, bhagtani2022overview}.
However, identifying information about the source of the audio is also important.
Different recording devices, editing software, and file compression methods leave clues about the nature of an audio signal.
It is possible to associate -- or attribute -- authentic audio signals to specific components used in their creation.
For example, Buchholz \etal identify microphones used to record audio signals by analyzing the signals' Fourier coefficients~\cite{buchholz2009microphone} with na\"{i}ve bayes, a support vector machine (SVM), logistic regression (LogReg), decision trees, and a k-nearest neighbors classifier (KNN).
Luo \etal focus specifically on smartphone attribution for audio recordings~\cite{luo_2018}.
They propose a feature set called a bank energy difference descriptor (BED) for analysis with a SVM.
Just as it is possible to attribute an authentic speech signal to a specific audio-capturing device, it is possible to identify speech synthesizers used to generate speech signals because each speech synthesizer leaves its own fingerprints in the audio signals it creates.
Borrelli \etal explore speech synthesizer attribution using a random forest, a linear SVM, and a non-linear SVM in both closed set and open set scenarios~\cite{borrelli_2021}.

In this work, we propose a deep learning method know as compact attribution transformer (CAT) for speech synthesizer attribution.
CAT is a convolutional transformer, meaning that it utilizes a series of convolutional layers before a transformer to train more efficiently and incorporate inductive biases into the network.
A convolutional transformer has achieved success in synthetic audio detection~\cite{bartusiak_2021_asilomar}, so now we explore its use for synthetic speech attribution.

Besides using deep learning, our approach differs from prior audio attribution work in that it analyzes spectrograms of audio signals.
We use spectrograms to leverage frequency information of speech signals.
Because spectrograms provide a structured version of a speech signal and group similar frequencies together, CAT can analyze different frequency ranges more explicitly to identify artifacts of various speech synthesizers.
Another novel aspect of our work is the use of t-distributed stochastic neighbor embedding (tSNE)~\cite{Maaten_2008} to discriminate between different unknown synthesizers. 
TSNE is an unsupervised learning method used for dimensionality reduction and visualizing high-dimensional data.
We use tSNE on the latent space of CAT to separate all known and unknown synthesizers.
Finally, we explore different loss functions to improve performance of the method.
We formulate the loss function used in CAT as poly-1 losses to tailor the loss function to this specific task.
We show that our proposed approach achieves success on this attribution task, even in an open set scenario.

%% file: part-2-method.tex
\section{Proposed Method}

Figure~\ref{fig:overview} shows an overview of our approach.
First, we convert all speech waveforms into normalized spectrograms~\cite{rs2010}.
Next, we use a series of convolutional layers (\ie a convolutional block) to extract feature maps from the spectrograms.
Then, a transformer analyzes the feature maps and produces a set of probabilities $P = \{p_i \mid 0 \leq i \leq N-1 \} $, where $N$ corresponds to the number of known synthesizers, and probability $p_i$ indicates the likelihood that a speech signal under analysis was generated by speech synthesizer $i$.
We use the probabilities in $P$ to determine which of the known speech synthesizers created the speech signal.
We also use a thresholding procedure on the probabilities to determine whether a speech signal was generated by a known or an unknown speech synthesizer.

\subsection{Spectrograms}

We convert speech waveforms into spectrograms by using a 32 ms Hanning window on the speech signals with a shift of 8 ms.
In other words, we use the Fast Fourier Transform (FFT) on blocks of the signals consisting of 512 sampled points with 128 points of overlap between consecutive blocks~\cite{rs2010}.
Note that all waveforms in our experimental dataset have a 16 kHz sampling rate.
Next, the FFT coefficients are converted to decibels and organized into 2-D arrays, where height corresponds to frequency bands and width corresponds to time (\ie length of the audio signal in terms of 32 ms time blocks).
We crop or pad each spectrogram to have dimensions 128x128 pixels to ensure that all inputs to the neural network are the same size. 
Finally, we normalize the spectrogram pixels to the range of values [0,1].
Normalized values enable machine learning models to learn more quickly because they are forced to focus on relative rather than absolute differences in input values. 
The normalized spectrograms of size 128x128 pixels are used as inputs to CAT.

\subsection{Compact Attribution Transformer (CAT)}

We propose a method called the compact attribution transformer (CAT) for speech synthesizer attribution.
It is a convolutional transformer that first extracts feature maps from an input spectrogram using convolutions.
Then, it uses an attention mechanism~\cite{vaswani_2017} to analyze the feature maps in order to attribute the audio signal under analysis to a speech synthesizer.
One significant benefit of our approach compared to other transformers is its size.
We design CAT to have 405 thousand parameters. 
There are a number of other transformers that have demonstrated success in image-based analyses, but they are significantly larger in size.
For example, the first transformer used on images -- known as Vision Transformer (ViT)~\cite{dosovitskiy_2021} -- has three variants of different sizes: ViT-Base has 86 million parameters; ViT-Large has 307 million parameters; and ViT-Huge has 632 million parameters.
ViT was adapted for spectrogram analysis to produce a model called Audio Spectrogram Transformer (AST)~\cite{gong_2021_ast}.
There are other variations of AST, such as Self-Supervised Audio Spectrogram Transformer (SSAST)~\cite{gong_2021_ssast}, that use self-supervised learning during training for audio classification.
Because AST and its variants (\eg SSAST) are based on ViT, they too have over 87 million parameters.
Patchout faSt Spectrogram Transformer (PaSST)~\cite{koutini_2021} is another transformer used to analyze spectrograms for audio classification.
It explicitly uses frequency information in its attention mechanism and has over 85 million parameters.
Because CAT uses significantly fewer parameters in comparison to these transformers, it trains faster and is easier to work with in scenarios with limited compute power. 

Another benefit of CAT is that it does not require an extremely large dataset for training~\cite{hassani_2021}.
This is important for our work because we investigate speech synthesizer attribution with a relatively small dataset containing only 17,000 samples (see Section~\ref{part-3-results} for details of the dataset).
Transformers lack some of the inductive biases that convolutions introduce, such as translational equivariance and locality~\cite{dosovitskiy_2021, hassani_2021}, and must learn these properties on their own.
However, this requires larger datasets that exhibit the properties.
Larger datasets are not always available for particular tasks (\eg medical tasks, our speech synthesizer attribution task), making it difficult to train transformers properly.
CAT explicitly introduces these biases into the network by using a convolutional block before the transformer.
The convolutional block also increases the parameter efficiency of CAT compared to non-convolutional transformers by using shared weights.
This enables CAT to train more efficiently with less data.

CAT is based on a network known as compact convolutional transformer~\cite{hassani_2021}.
Our CAT network uses two convolutional layers in the convolutional block. 
Each convolutional layer uses a 3x3 kernel, ReLU activation functions, max pooling, and positional embedding. 
The transformer encoder contains two transformer layers that use layer normalization, two attention heads, GeLU activation functions, stochastic depth, and sequence pooling~\cite{hassani_2021}.
The final dense layer of the network uses a softmax function to create the set of output probabilities $P$.
We train CAT for 100 epochs with a patience of 10 epochs and use the AdamW optimizer~\cite{loshchilov_2019} with an initial learning rate and weight decay of $10^{-4}$.
The batch size in our experiments is 128.

\subsection{Loss Functions}

We investigate the use of three different loss functions in this work.
First, we use a standard cross entropy (CE) loss~\cite{bishop_2006}.
Then, we use a poly-1 cross entropy loss (poly-1-CE) and a poly-1 focal loss (poly-1-FL)~\cite{leng_2022}.
Poly-1 losses approximate loss functions via Taylor expansion.
This allows loss functions to be designed as linear combinations of polynomial functions with easily adjusted polynomial bases. 
The polynomial bases can be customized for a specific dataset or specific task, enabling better performance with the same dataset and model architecture.

Let us begin with the definition of cross entropy (CE) loss: $L_{CE} = - log(p_i).$ The poly-1 cross entropy loss is then: $ L_{poly-1-CE} = L_{CE} + \epsilon(1 - p_i).$ Similarly, we can formulate the poly-1 focal loss. Let us begin with the definition of focal loss (FL): $ L_{FL} = - (1 - p_i)^{\gamma}log(p_i), $ where $\gamma$ serves as a modulating factor.
It reduces the loss contribution from ``easy" data samples by extending the range of probabilities that contribute to a low loss value. Note that when $\gamma = 0$, focal loss is equivalent to cross entropy loss (\ie $L_{FL} = L_{CE}$). Then, the poly-1 focal loss is: $ L_{poly-1-FL} = L_{FL} + \epsilon(1 - p_i)^{\gamma + 1}. $

By constructing the loss function with this poly-1 formulation, we can adapt the loss function for our specific dataset by changing $\epsilon$.
We perform a gridsearch to find the best value of $\epsilon$ for our task.
Experiments indicate that $\epsilon = 3.3$ yields the best results with poly-1 cross entropy loss, and $\epsilon = 3$ yields the best results with poly-1 focal loss.
Note that we use $\gamma = 2$ with poly-1 focal loss.
More details of these experiments are included in Section~\ref{part-3-results}.

\subsection{Identifying Unknown Synthesizers}

In addition to attributing synthesized speech signals to known synthesizers, we also determine if a speech signal is created by an unknown synthesizer.
To do so, we implement a thresholding procedure on the probabilities in $P$ produced by CAT.
We define a threshold $T$ that determines if CAT is confident in its attribution or not.
Let $p_m$ represent the maximum probability in $P$ $\left( \textrm{\ie} p_m = max(P) \right)$.
If $p_m > T$, then CAT is very confident that it can attribute the speech signal under analysis to the known synthesizer $i$.
If none of the probabilities in $P$ are greater than the threshold $T$, CAT cannot detect any fingerprints from any of the known synthesizers.
Instead, the signal must be created by an unknown synthesizer.
In this case, it is assigned to the category containing all unknown synthesizers $U$.
In practice, any threshold $T$ can be selected depending on the application and dataset.

\begin{table*}[ht]
\normalsize
{\rowcolors{3}{white!10}{lightskyblue!10}
\begin{center}
    \caption{Dataset used in our experiments.}
    {\renewcommand{\arraystretch}{1.05} 
    \begin{tabular}{ |C{2.8cm}|C{2.5cm}|C{2.5cm}|C{2.1cm}|C{3.0cm}|C{1.6cm}| }
    \hline
    \rowcolor{lightskyblue!40} \multicolumn{6}{|c|}{\textbf{SemaFor Audio Model Attribution Dataset}} \\
    \hline
    \rowcolor{lightskyblue!15!}
    Speech Synthesizer & Training Samples & Testing Samples & Total Samples & Speakers & Type\\
    \hline
    FastPitch       & 1,000 &  1,500 &  2,500 & 1 from LJSpeech & Known\\
    FastSpeech2     &   500 &    300 &    800 & 1 from LJSpeech & Known\\
    Glow-TTS        & 1,000 &  1,500 &  2,500 & 1 from LJSpeech & Known\\
    gTTS            & 1,000 &  1,500 &  2,500 & 4 unknown & Known\\
    Tacotron        & 1,000 &  1,500 &  2,500 & 10 from LibriSpeech & Known\\
    Tacotron 2      &   500 &    300 &    800 & 1 from LJSpeech & Known\\
    TalkNet         & 1,000 &  1,500 &  2,500 & 1 from LJSpeech & Known\\
    Riva            & 1,000 &  1,000 &  2,000 & 1 from LJSpeech & Known\\
    Mixer-TTS       &     0 &    300 &    300 & 1 from LJSpeech & Uknown\\
    SpeedySpeech    &     0 &    300 &    300 & 1 from LJSpeech & Uknown\\
    VITS            &     0 &    300 &    300 & 10 from VCTK & Uknown\\
    \hline
    Total           & 7,000 & 10,000 & 17,000 & 25 & \\
    \hline
    \end{tabular}
    \vspace{-.35cm}
    }
    \label{tab:dataset}
\end{center}
}
\end{table*}

\begin{table}[ht]
    \begin{center}
    \caption{Best hyperparameters of all methods considered.}
    \resizebox{0.99\columnwidth}{!}{
        {\renewcommand{\arraystretch}{1.015} 
        \begin{tabular}{lccc}
	        \toprule
            \textbf{Method} &  \textbf{Hyperparameter} & \textbf{Value}\\
            \midrule
            QDA                         & \texttt{reg\_param} & \texttt{0.0} \\
            \hline
            GP                          & \texttt{optimizer} & \texttt{fmin\_l\_bfgs\_b} \\
            \hline
            AdaBoost                    & \texttt{n\_estimators} & \texttt{50} \\
            \hline
                                        & \texttt{n\_neighbors} & \texttt{2} \\
                                        & \texttt{weights} & \texttt{distance} \\
            KNN                         & \texttt{algorithm} & \texttt{auto} \\
                                        & \texttt{leaf\_size} & \texttt{10} \\
                                        & \texttt{p} & \texttt{1} \\
            \hline
            Na\"{i}ve Bayes             & \texttt{var\_smoothing} & \texttt{1e-7} \\
            \hline
                                        & \texttt{criterion} & \texttt{gini} \\
            Decision Tree               & \texttt{splitter} & \texttt{best} \\
                                        & \texttt{min\_samples\_split} & \texttt{10} \\
                                        & \texttt{class\_weight} & \texttt{balanced} \\
            \hline
                                        & \texttt{hidden\_layer\_sizes} & \texttt{(1500, 1500)} \\
                                        & \texttt{activation} & \texttt{logistic} \\
                                        & \texttt{batch\_size} & \texttt{200} \\
                                        & \texttt{epochs} & \texttt{200} \\
            MLP                         & \texttt{patience} & \texttt{10} \\
                                        & \texttt{early\_stopping} & \texttt{True} \\
                                        & \texttt{optimizer} & \texttt{adam} \\
                                        & \texttt{learning\_rate\_init} & \texttt{0.0001} \\
                                        & \texttt{loss\_function} & \texttt{CE} \\
            \hline
                                        & \texttt{n\_estimators} & \texttt{500} \\
            Random Forest               & \texttt{min\_samples\_split} & \texttt{5} \\
                                        & \texttt{class\_weight} & \texttt{balanced} \\
            \hline
                                        & \texttt{C} & \texttt{1} \\
                                        & \texttt{kernel} & \texttt{poly} \\
                                        & \texttt{degree} & \texttt{2} \\
            Non-Linear                  & \texttt{gamma} & \texttt{scale} \\
            SVM                         & \texttt{shrinking} & \texttt{True} \\
                                        & \texttt{class\_weight} & \texttt{None} \\
                                        & \texttt{probability} & \texttt{True} \\
                                        & \texttt{max\_iter} & \texttt{100} \\
            \hline
                                        & \texttt{C} & \texttt{0.1} \\
                                        & \texttt{kernel} & \texttt{linear} \\
            Linear                      & \texttt{shrinking} & \texttt{True} \\
            SVM                         & \texttt{class\_weight} & \texttt{None} \\
                                        & \texttt{probability} & \texttt{True} \\
                                        & \texttt{max\_iter} & \texttt{100} \\
            \hline
                                        & \texttt{penalty} & \texttt{L1} \\
                                        & \texttt{C} & \texttt{10} \\
            LogReg                      & \texttt{class\_weight} & \texttt{balanced} \\
                                        & \texttt{solver} & \texttt{liblinear} \\
                                        & \texttt{max\_iter} & \texttt{500} \\
                                        & \texttt{l1\_ratio} & \texttt{0.75} \\
            \hline
                                        & \texttt{batch\_size} & \texttt{128} \\
                                        & \texttt{epochs} & \texttt{100} \\
            CNN                         & \texttt{patience} & \texttt{10} \\
                                        & \texttt{optimizer} & \texttt{adam} \\
                                        & \texttt{learning\_rate\_init} & \texttt{0.001} \\
                                        & \texttt{loss\_function} & \texttt{CE} \\
            \hline
                                        & \texttt{batch\_size} & \texttt{128} \\
                                        & \texttt{epochs} & \texttt{100} \\
                                        & \texttt{patience} & \texttt{10} \\
            CAT                         & \texttt{optimizer} & \texttt{adamW} \\
                                        & \texttt{learning\_rate\_init} & \texttt{0.001} \\
                                        & \texttt{weight\_decay\_init} & \texttt{0.001} \\
                                        & \texttt{loss\_function} & \texttt{poly-1-CE} \\
                                        & \texttt{$\epsilon$} & \texttt{3.3} \\
	        \bottomrule
	        \vspace{-1cm}
        \end{tabular}}
        }
        \label{tab:best-hyperparameters}
  \end{center}	
\end{table}

This thresholding procedure enables attribution to $N+1$ classes (\ie to one of the known synthesizers or to the unknown category $U$).
However, it does not allow for discrimination between different unknown synthesizers within $U$.
To attribute a synthesized signal to a specific unknown synthesizer, we utilize t-distributed stochastic neighbor embedding (tSNE)~\cite{Maaten_2008} on the latent space of CAT.
First, we embed all speech signals in the open set into the latent space of CAT by running them through the model.
Next, we use tSNE on the output of CAT's transformer block, which is the layer directly before the final dense layer that maps the latent space of CAT to a lower-dimensional output vector of length $N$ (\ie the number of known synthesizers).
TSNE projects the high-dimensional latent space of CAT onto two dimensions, allowing visualization of CAT's internal representation of all of the speech signals in the open set.
Then, we examine the tSNE visualization to understand how CAT separates different synthesizers in its latent space.
Our results indicate that distinct clusters form in the tSNE visualization and that each cluster corresponds to a different synthesizer.
Even the different unknown synthesizers have distinct clusters in the latent space.
Thus, analyzing the tSNE visualization provides more details about the nature of synthesized speech signals. 
We present our findings in Section ~\ref{part-3-results}. 
In our experiments, we use a perplexity of 50 and 1,500 iterations to fit tSNE to CAT's latent space.

%% file: part-3-results.tex
\section{Results}\label{part-3-results}

\subsection{Dataset}

To validate our approach, we use a dataset known as the SemaFor Audio Model Attribution Dataset.
It was presented in the Semantic Forensics (SemaFor) program~\cite{semafor} organized by the Defense Advanced Research Projects Agency (DARPA). 
The dataset is based on a dataset from the 2022 IEEE Signal Processing Cup (SP Cup)~\cite{sp_cup_2022}, which is a competition related to synthetic speech attribution.
The dataset from the SP Cup was augmented with another speech synthesizer (known as Riva~\cite{riva}) and re-arranged to create the final version of the SemaFor Audio Model Attribution Dataset.

There are a total of 17,000 synthesized speech signals in the dataset from 11 different speech synthesizers.
Speech signals from eight synthesizers are considered to be ``known" (\ie part of the closed set).
Speech signals from the remaining three synthesizers are considered to be ``unknown" (\ie part of the open set).
The speech synthesizers in the closed set are FastPitch\cite{fastpitch_2021}, FastSpeech2~\cite{fastspeech2_2021}, Glow-TTS~\cite{glowtts_2020}, gTTS~\cite{gtts}, Tacotron~\cite{tacotron_2017}, Tacotron 2~\cite{tacotron2_2018}, TalkNet~\cite{talknet_2020}, and Riva~\cite{riva}.
The speech synthesizers in the open set are Mixer-TTS~\cite{mixertts_2022}, SpeedySpeech~\cite{speedyspeech_2020}, and VITS~\cite{vits_2021}.
The speech synthesizers use various deep learning methods, including transformers~\cite{vaswani_2017}, generative flow models~\cite{hoogeboom_2019, kingma_2018}, convolutional neural networks (CNNs)~\cite{wavenet_2016}, long short-term memory networks (LSTMs)~\cite{hochreiter_1997, sutskever_2014}, multilayer perceptron networks (MLPs)~\cite{tolstikhin_2021}, generative adversarial networks (GANs)~\cite{goodfellow_2014, kumar_2019, kong_2020}, and variational autoencoders (VAEs)~\cite{kingma_2014} to generate new speech signals. 
All signals have a sampling rate of 16 kHz.

Table~\ref{tab:dataset} summarizes the details of the dataset. 
We utilize the official dataset split according to the SemaFor program for training and testing our approach.
Notice that there are an unequal number of samples associated with each speech synthesizer.
For example, Tacotron 2 and FastSpeech2 only have 500 training samples, but the other speech synthesizers in the closed set have 1,000 training samples.
Similarly, these two synthesizers have only 300 testing samples, while some other synthesizers have 1,500 testing samples.
This imbalance presents a challenge in training an attribution model because there are fewer data samples that can be used to learn to recognize certain speech synthesizers.
One other challenge with this dataset is that not all speech synthesizers use the same speaker (\ie simulate the same voice).
Most of the speech synthesizers replicate a speaker from the LJSpeech dataset~\cite{ljspeech_2017}, but some synthesizers use speakers from the LibriSpeech dataset~\cite{librispeech_2015}, the VCTK dataset~\cite{vctk_2012}, and an unknown dataset used by Google.
Because there are different speakers in the dataset, an audio attribution model must learn the characteristics of a speech synthesizer regardless of the speaker.

\subsection{Evaluation Process}

We report accuracy, weighted precision, weighted recall, and weighted F-1~\cite{tharwat_2021} of all methods evaluated. 
Weighted metrics are computed with a weighted average of each metric obtained on each of the classes in the dataset, where weights reflect the dataset class imbalance.
Recall that all evaluated methods produce a set $P$ of $N$ probabilities, where each probability represents the likelihood that a known synthesizer was used to generate a speech signal under analysis.
In a closed set scenario, the final attribution output is the synthesizer that corresponds to the maximum probability $p_m \in P$. 
All metrics are computed based on only the $N$ known synthesizers.
In an open set scenario, the thresholding procedure produces the final attribution output, which could either be one of the known synthesizers or an unknown class $U$.
In this case, all metrics are computed based on the $N+1$ classes.

\subsection{Baselines and Comparison Methods}

To validate our approach, we compare it against theoretical baselines, classical machine learning methods, and deep learning methods.
We refer to our theoretical baselines as Baseline-Minority and Baseline-Majority.
These baselines represent theoretical classifiers that only predict one synthesizer.
In other words, every time a new speech signal is presented to the theoretical classifiers, they always predict the same attribution output.
In the case of Baseline-Minority, the output is always the minority class (\ie the dataset class that corresponds to the fewest samples in the training set).
In the case of Baseline-Majority, the output is always the majority class (\ie the dataset class that corresponds to the most samples in the training set).
In the case where multiple classes have the same number of data samples, one of the tied classes is chosen to represent the minority/majority class.
Because these baselines are computed from the experimental dataset, they establish lower bounds of the evaluation metrics to indicate whether a classifier actually performs well.

In addition to these baselines, we investigate several classical machine learning methods: quadratic discriminant analysis (QDA); gaussian process (GP); AdaBoost; k-nearest neighbors (KNN); na\"{i}ve bayes; decision tree; random forest; non-linear and linear support vector machines (SVMs); and logistic regression (LogReg). 
To construct the inputs to these methods, the normalized spectrograms of size 128x128 pixels are row concatenated to produced flattened, 1-D arrays of length 16,384.
Each of these methods has various hyperparameters that can affect their performance, so we execute a gridsearch to determine the best set of hyperparameters for each method.
Table~\ref{tab:best-hyperparameters} specifies the best hyperparameters for all methods included in our analysis.

Finally, we explore two deep learning methods in addition to our proposed CAT solution: a multi-layer perceptron network (MLP) and a convolutional neural network (CNN).
The MLP operates on the 1-D version of the spectrograms, while the CNN operates on the original 2-D format (\ie the same format analyzed by CAT).
The MLP consists of two hidden layers of 1,500 nodes with logistic activation functions and one output layer that consists of 8 nodes (equal to the number of known synthesizers in our experimental dataset).
We train the MLP for 200 epochs with a patience of 10 epochs using the Adam optimizer~\cite{adam} with an initial learning rate of $10^{-4}$.
The batch size is 200 in our experiments.
The CNN consists of two convolutional layers followed by two dense layers.
The convolutional layers use a 3x3 kernel and ReLU activation functions. 
The first dense layer contains 128 nodes, and the second dense layer (\ie the output layer) contains 8 nodes.
The CNN uses max pooling and dropout to regularize the network.
We train the CNN for 100 epochs with a patience of 10 epochs using a batch size of 128.
The CNN uses the Adam optimizer~\cite{adam} with an initial learning rate of $10^{-3}$.
Both the MLP and the CNN use a softmax activation function on the outputs of the final dense layer to create a set of output probabilities $P$.
They also both use cross entropy as a loss function.

\subsection{Experimental Results}

Table~\ref{tab:results-losses} shows results with CAT using different loss functions.
The baseline loss function is cross entropy $L_{CE}$, which achieves approximately 90\% accuracy, weighted precision, weighted recall, and weighted F-1 on the closed set.
Both poly-1 loss formulations improve attribution results compared to this baseline.
Poly-1 focal loss ($L_{poly-1-FL}$) improves all results, especially in terms of accuracy and weighted recall.
Poly-1 cross entropy loss ($L_{poly-1-CE}$) improves results even further.
All metrics increase by roughly 2\% with $L_{poly-1-CE}$, resulting in 92.53\% accuracy and 91.27\% F-1.
Because poly-1 cross entropy achieves the greatest results overall, CAT experiments reported in the rest of this paper use $L_{poly-1-CE}$.

\begin{table}[ht]
    \begin{center}
    \caption{CAT results on closed set with different losses.}
    \resizebox{1.0\columnwidth}{!}{
        {\renewcommand{\arraystretch}{1.2} 
        \begin{tabular}{lcccc}
	        \toprule
            \textbf{Loss Function} &  \textbf{Accuracy} & \textbf{Precision} & \textbf{Recall} & \textbf{F-1}\\
            \midrule
            $L_{CE}$            & 90.12\%           & 89.05\%           & 90.12\%           & 89.45\% \\
            $L_{poly-1-CE}$     & \textbf{92.53\%}  & \textbf{90.37\%}  & \textbf{92.53\%}  & \textbf{91.27\%}\\
            $L_{poly-1-FL}$     & 92.02\%           & 89.65\%           & 92.02\%           & 90.67\%\\
	        \bottomrule
        \end{tabular}}
        }
        \label{tab:results-losses}
  \end{center}	
\end{table}

\begin{figure}[h]
    \centering
    \includegraphics[width=.896\columnwidth]{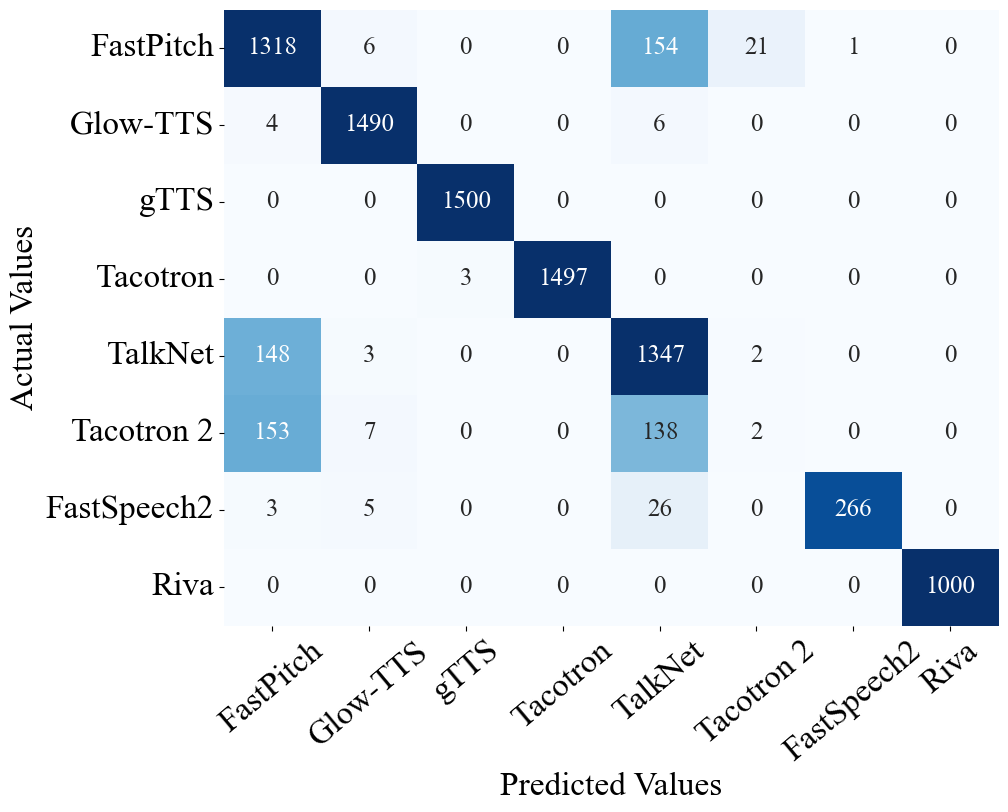}
    \caption{Confusion matrix of CAT results with poly-1-CE loss.}
    \label{fig:confusion}
    \vspace{-.5cm}
\end{figure}

Table~\ref{tab:results-known} shows results of all methods on the closed set. 
All methods outperform the theoretical baselines and demonstrate that they achieve greater attribution success than minority class and majority class classifiers.
However, some methods cannot even achieve 50\% accuracy on this attribution task.
For example, all metrics obtained with QDA, GP, and AdaBoost are less than 35\%.
KNN barely passes the 50\% mark. 
Thus, not all methods are suited for synthesizer attribution.
Decision trees achieve better success on this task, obtaining approximately 70\% accuracy.
Using an ensemble of decision trees (\ie a random forest) improves results even further to $\sim$80\% accuracy.
SVMs achieve better attribution performance, with both linear and non-linear SVMs achieving roughly 81\% accuracy on this task.
LogReg achieves the best performance of the classical machine learning methods ($\sim$90\%).
However, two deep learning methods achieve better performance.
Both CNN and CAT achieve 90\% or higher for all metrics considered.
Our CAT with poly-1 cross entropy loss outperforms all methods for all metrics, achieving $\sim$93\% accuracy and $\sim$91\% F-1.
Figure~\ref{fig:confusion} shows more details about the performance of CAT on the closed set.
This confusion matrix indicates that CAT can attribute speech signals to most of the known synthesizers (7 out of 8) very well. 
CAT struggles with just one synthesizer: Tacotron 2. 
Tacotron 2 is one of the synthesizers with the fewest training and testing samples.
It is more difficult for an attribution method to learn to identify a synthesizer from fewer samples, which is one factor contributing to the challenge in identifying Tacotron 2 speech signals. 

\begin{table}[h]
    \begin{center}
    \caption{Results of all methods on closed set.}
    \resizebox{1.0\columnwidth}{!}{
        {\renewcommand{\arraystretch}{1.1} 
        \begin{tabular}{lcccc}
	        \toprule
            \textbf{Method} &  \textbf{Accuracy} & \textbf{Precision} & \textbf{Recall} & \textbf{F-1}\\
            \midrule
            Baseline-Minority           & 3.30\% & 0.11\% & 3.30\% & 0.21\% \\
            Baseline-Majority           & 16.48\% & 2.72\% & 16.48\% & 4.67\% \\
            QDA                         & 19.34\% & 16.82\% & 19.34\% & 11.75\% \\
            GP                          & 21.62\% & 68.80\% & 21.62\% & 12.82\% \\
            AdaBoost                    & 33.58\% & 32.71\% & 33.58\% & 23.69\% \\
            KNN                         & 52.22\% & 56.63\% & 52.22\% & 49.83\% \\
            Na\"{i}ve Bayes             & 68.14\% & 70.95\% & 68.14\% & 69.08\% \\
            Decision Tree               & 69.02\% & 71.08\% & 69.02\% & 69.93\% \\
            MLP                         & 78.91\% & 77.06\% & 78.91\% & 77.68\% \\
            Random Forest               & 81.04\% & 79.90\% & 81.04\% & 79.10\% \\
            Non-Linear SVM              & 81.13\% & 81.35\% & 81.13\% & 81.05\% \\
            Linear SVM                  & 81.57\% & 80.99\% & 81.57\% & 81.22\% \\
            LogReg                      & 90.68\% & 88.29\% & 90.68\% & 89.43\% \\
            CNN                         & 91.99\% & 90.21\% & 91.99\% & 90.88\% \\
            CAT                         & \textbf{92.53\%} & \textbf{90.37\%} & \textbf{92.53\%} & \textbf{91.27\%} \\
	        \bottomrule
        \end{tabular}}
        }
        \label{tab:results-known}
  \end{center}	
\end{table}


\begin{table}[h]
    \begin{center}
    \caption{Results of all methods on open set.}
    \resizebox{.99\columnwidth}{!}{
        {\renewcommand{\arraystretch}{1.1} 
        \begin{tabular}{lcccc}
	        \toprule
            \textbf{Method} &  \textbf{Accuracy} & \textbf{Precision} & \textbf{Recall} & \textbf{F-1}\\
            \midrule
            Baseline-Minority           & 3.00\% & 0.09\% & 3.00\% & 0.17\% \\
            Baseline-Majority           & 15.00\% & 2.25\% & 15.00\% & 3.91\% \\
            QDA                         & 17.60\% & 14.45\% & 17.60\% & 9.86\% \\
            GP                          & 9.00\% & 0.81\% & 9.00\% & 1.49\% \\
            AdaBoost                    & 17.51\% & 10.84\% & 17.51\% & 12.80\% \\
            KNN                         & 47.52\% & 46.89\% & 47.52\% & 42.77\% \\
            Na\"{i}ve Bayes             & 62.01\% & 59.58\% & 62.01\% & 59.95\% \\
            Decision Tree               & 62.66\% & 60.01\% & 62.66\% & 60.77\% \\
            MLP                         & 55.97\% & 57.03\% & 55.97\% & 53.36\% \\
            Random Forest               & 47.74\% & 80.92\% & 47.74\% & 46.29\% \\
            Non-Linear SVM              & 65.41\% & 73.24\% & 65.41\% & 66.41\% \\
            Linear SVM                  & 68.47\% & 71.78\% & 68.47\% & 68.80\% \\
            LogReg                      & 83.85\% & 81.44\$ & 83.85\% & 81.62\% \\
            CNN                         & 83.56\% & 77.26\% & 83.56\% & 79.67\% \\
            CAT                         & \textbf{84.10\%} & \textbf{82.37\%} & \textbf{84.10\%} & \textbf{83.00\%} \\
	        \bottomrule
        \end{tabular}}
        }
        \label{tab:results-unknown}
  \end{center}
  \vspace{-.5cm}
\end{table}

Table~\ref{tab:results-unknown} shows results on the open set using the thresholding method.
In general, attribution results drop for all approaches in the open set scenario.
One method's performance (GP) even drops below Baseline-Majority.
Other methods, such as QDA and AdaBoost, barely beat Baseline-Majority based on some metrics.
Most methods' metrics drop to under 80\%.
Only three methods achieve metrics greater than 80\%: LogReg, CNN, and CAT.
This drop in performance indicates the difficulty of synthesizer attribution in an open set scenario.
However, CAT is still able to attribute synthesized speech with over 84\% accuracy and 83\% F-1, achieving the highest attribution results overall.
Thus, results indicate that CAT with poly-1 cross entropy loss is the best method for synthesized speech attribution in both closed set and open set scenarios.


\begin{figure}[ht]
    \centering
    \includegraphics[width=.99\columnwidth]{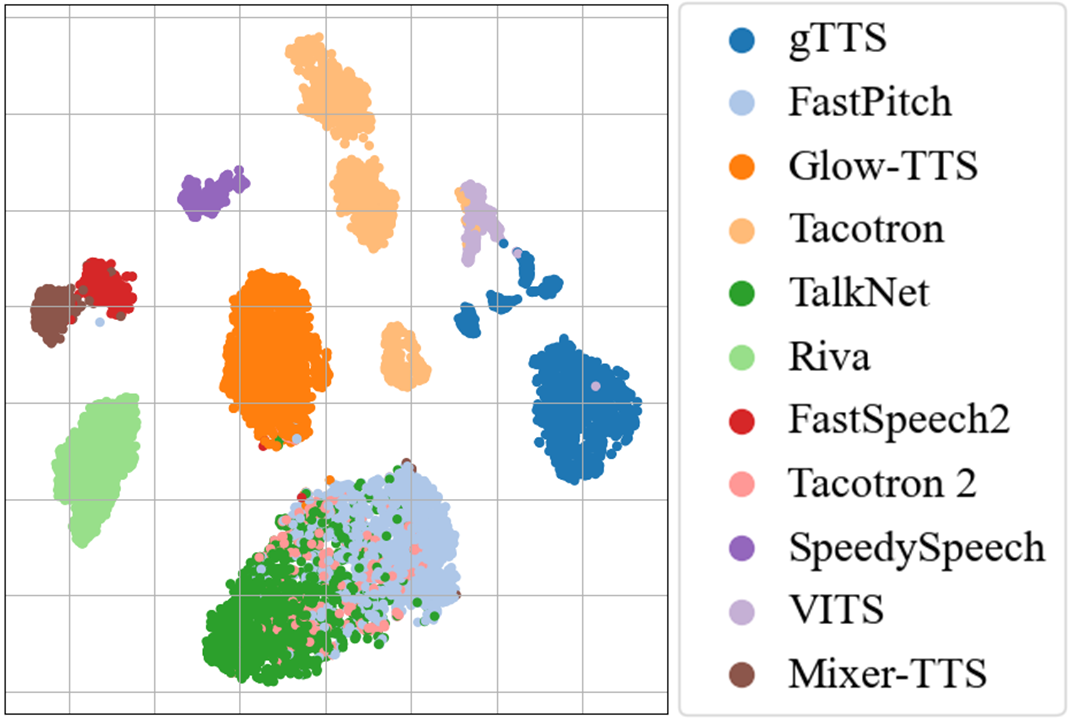}
    \caption{TSNE plot of CAT's latent space on open set. Mixer-TTS, SpeedySpeech, and VITS are unknown synthesizers.}
    \label{fig:tsne}
\end{figure}


Figure~\ref{fig:tsne} shows CAT's latent space representation of all speech signals in the open set.
We explore the tSNE plot to better understand the attribution performance of CAT.
It shows distinct clusters for most of the synthesizers in our experimental dataset.
For example, gTTS, Glow-TTS, Tacotron, Riva, FastSpeech2, SpeechSpeech, VITS, and Mixer-TTS all have their own clusters within the latent space. 
This separability indicates that CAT can successfully identify unique fingerprints of each of these synthesizers and use them for attribution.
Recall that Mixer-TTS, SpeedySpeech, and VITS are all unknown synthesizers.
Thus, CAT can actually discriminate between different unknown synthesizers, even though they have never been presented to the network before.
This ability can provide further information to forensic analysts and help them in real-world scenarios as new synthesizers are invented.
TSNE also provides further details about CAT's vulnerabilities.
Although most synthesizers are separated into distinct clusters in the tSNE plot, three synthesizers have less separability: TalkNet, Tacotron 2, and FastPitch.
The tSNE plot indicates that TalkNet and FastPitch have more separability from each other, but Tacotron 2 has extensive overlap with TalkNet and FastPitch.
Thus, tSNE confirms results summarized by the confusion matrix in Figure~\ref{fig:confusion}.
Figure~\ref{fig:confusion} indicates that CAT is not able to attribute Tacotron 2 speech signals, and Figure~\ref{fig:tsne} provides evidence that CAT cannot separate those signals correctly.
To try to mediate this issue, we explored weighted loss functions that account for the dataset's class imbalance.
We also investigated methods that could discriminate between Tacotron 2 speech signals and all other synthesizers in a one-vs-all fashion.
However, these initial investigations were unsuccessful at attributing Tacotron 2 signals.
In future work, we seek to overcome this challenge.

%% file: part-4-conclusion.tex
\section{Conclusion}\label{part-4-conclusion}

In this paper, we attribute speech signals to speech synthesizers in both closed set and open set scenarios.
We analyze speech signals in the form of spectrograms with a compact attribution transformer (CAT) and demonstrate that our approach achieves success on this speech synthesizer attribution task.
Furthermore, we show that poly-1 loss formulations improve results and that analyzing the latent space of CAT with tSNE can discriminate between different unknown synthesizers.
Future work will extend this analysis to more speech synthesizers and increase separability in the latent space. 